# Physics of complex systems: Discovery in the age of Gödel

Dragutin T. Mihailović*,(1), Darko Kapor*,(1), Siniša Crvenković*,(2) and Anja Mihailovic**
University of Novi Sad, Novi Sad, Serbia
*Faculty of Sciences, **Faculty of Technical Sciences
(1) Department of Physics, (2) Department of Mathematics and Informatics

## Chapter 1

**Prolegomena**

From the perspective of the physics of complex systems (1) we deal with the current state of modern physics including the crisis in physics demonstrated through its epistemological, psychological, economical as well as the social context; (2) considering the strength of the Gödel's Incompleteness Theorems we point out the following open questions in physics: (i) the limits of the precision of certainty, (ii) the limits of making decisions on information and (iii) the limitations of reasoning that sometimes affect progress, (3) since further advances will necessarily require synergy between physics and seemingly distinct fields - mathematics, information science, chemistry, biology, medicine, psychology, and art. We illustrate this relation by providing examples based on our research. Point (1) is discussed in Ch1 while Ch2-Ch5 encompasses the point (2) and point (3) is covered in Ch6-Ch10.

**1.2 Physics: A crisis that has been lasting for a century! Is that really so?**

*When we have found all the mysteries and lost all the meaning, we
will be alone, on an empty shore.*

Tom Stoppard, *Arcadia*

We will now discuss a crisis and our perspective on its definition and origins. Many articles on the crisis in science have been published starting from the pioneering paper written by Freistadt (1953) which is probably the first paper specifying this problem. Nonetheless, the clear identification of its origins remains elusive to most commentators (Saltelli and Funtowicz 2017). Due to the increasing public impact of the crisis on trust in institutions, authors argued that: (i) the crisis in science (hence physics) exists and encompasses positions and social functions of science (crisis context); (ii) the mainstream interpretation of its causes is not sufficient, and has to be complemented with insights provided by some researchers who predicted the current dilemma (a lack of root causes); (iii) the profound transformations of society and the impact of science on society are, without doubt, induced by this crisis. Scientists have also contributed to its creation and had a great influence on preserving the status quo (science vs stake context); (iv) there are some social mechanisms that can be applied to improve this situation including important changes in behavior and social activity of the scientific community (social context) (emphasis added by the authors). The crisis and its consequences are clearly reflected in awarding the Nobel Prize.

Fortunato et al (2014) point out that the time between publishing a discovery and receiving the Nobel Prize has become longer. This trend is the least present in physiology or medicine and the most present in physics. Prior to 1940, only 11 percent of prizes in physics, 15 percent of prizes in chemistry, and 24 percent of prizes in physiology or medicine were awarded for research more than 20 years old. Since 1985, those percentages have risen to 60 percent, 52 percent, and 45 percent, respectively.

"The crisis *of* physics" could imply that everything is affected and that all results are questionable. According to our opinion this is not the issue, and we will just state that a crisis exists *within* physics, or *in* physics. What is the crisis in physics? We perceive it as the absence of a fundamental discovery or, more precisely, the crisis in physics arises and lasts until a new fundamental discovery is made. This expectation is not similar to the one in Samuel Beckett's tragicomedy *Waiting for Godot*, the expectation of a person that will never appear. Many modern concepts about the crisis in physics envisage ending identical to the end of this tragicomedy; we do not see ending, but rather transition. From our standpoint, a fundamental discovery is a discovery made by scientists who possess exceptional abilities of thinking and revolutionize physics. We will label such progress as *vertical progress,* and it does not matter if it involves micro, macro, or mega physical world. Other discoveries in physics, the confirmation of already existing hypotheses or experimental results offering providing or needing explanations, will be entitled *horizontal progress*. We must admit that our definition of a fundamental discovery is not complete, since it assumes a large amount of horizontal progress that precedes the discovery. For this reason, we are not confident whether the experimental proof of the expanding universe and the quark hypothesis should be treated as fundamental.

In our opinion, the three most recent fundamental discoveries in physics are: (1) Quantum Theory (Max Planck, 1900) that introduced an individual unit of energy or *quant*; the sources are the two outstanding communications of Planck to the Berlin Academy (19-X-1900, extraordinary fit of the radiation formula, and 14-XII-1900, statistical justification by introduction of the discrete energy elements) collected in paper by Boya (2004). (2) Theory of Special 1905 was his *annus mirabilis* year when he published three papers in *Annalen der Physik* (1905a-1905c). These papers in which he revolutionized the concepts of space, time, mass, and energy made major contributions to the foundation of modern physics. Let us note that the contraction of a moving body in the direction of its motion was proposed independently by Hendrik Lorentz and George FitzGerald in 1892 to account for the null result of the Michelson–Morley experiment. (3) Theory of General Relativity (Albert Einstein, 1915) which is concerned with macroscopic behavior, and describes large-scale physical phenomena. Sauer (2004) provided the first comprehensive overview of the final version of the general theory of relativity published by Einstein in 1916 after several expositions of preliminary versions and their latest revisions in November 1915. These theories treat a microworld (molecules, atoms, and elementary particles), a macro world (our planet and people who represent macro objects) and a mega world or universe (mega objects such as stars, planets etc.). Here it should be stressed that almost all facts relevant to fundamental discoveries were previously available to scientists. Planck was attempting to explain his extraordinary fit. Due to Lobachevsky, Riemann and Minkowski, the non-Euclidean geometry was already well known before the General Theory of Relativity. Prior to the Theory of Special Relativity, physicists were familiar with: (1) *Lorentz force* (James Maxwell, and Hendrik Lorentz in 1895 (Tombe 2012)). (2) *Lorentz transformations.* These transformations which are considered constitutive for the Special Relativity Theory were invented by Woldemar Voigt in 1887 while their derivation is described in Heras (2017). It is interesting that Lorentz admitted that his space-time transformations were first introduced by Voigt (Kox 2008; Heras 2017)). (3) (*Einstein's*) *equation which relates mass to energy*. Henri Poincare stated that if one required that the momentum of any particle present in an electromagnetic field plus the momentum of the field itself should be conserved together, then Poynting's theorem (Poynting 1884) predicted that the field acts as a "fictitious fluid" with mass such that. Unfortunately, Poincare failed to connect with the mass of any real body. The only difference between the above-mentioned scientists and Einstein is the way they *interpreted* available facts. His interpretation made his discovery fundamental which he did by relating the facts in a manner which raised physics to a higher level. All three fundamental discoveries occurred within the first fifteen years of the twentieth century, so we are still waiting. During this expectation many things have happened which have brought us closer or distant us from the next such event.

Apart from fundamental discoveries, other important scientific advances have emerged in physics from 1850 to 2020. They include experimental discoveries, theoretical proposals that have been confirmed experimentally and theories that have remarkably influenced modern physics. We will mention some of them without years and the names of scientists which are well-known facts.

*(1850-1899)*.(1) A Treatise on Electricity and Magnetism; (2) Kinetic theory of gases; (3) Black body radiation; (4) Maxwell's equations; (5) Entropy; (6) A Dynamical Theory of the Electromagnetic Field (electromagnetic radiation); (7) Statistical mechanics (Boltzmann equation, 1872); (8) Stefan radiation law; (9) Boltzmann derivation of Second law of thermodynamics; (10) Michelson–Morley experiment; (11) Electromagnetic waves; (12) Rydberg formula; (13) Lorentz-FitzGerald contraction; (14) Wien's displacement law for black-body radiation; (15) X-rays; (16); Radioactivity; (17) Zeeman effect (18) Electron discovered.

*(1900-1950)*. (19) Formula for black-body radiation - the quanta solution to radiation ultraviolet catastrophe; (20) Thomson's plum pudding model of the atom; Special relativity, the introduction of a photon to explain the photoelectric effect, Brownian motion, Mass–energy equivalence; (21) Discovery of the atomic nucleus (Rutherford model); (22) Schwarzschild metric modeling gravity outside a large sphere; (23) Light bending confirmed - evidence for general relativity; (24) Kaluza–Klein theory proposing unification of gravity and electromagnetism; (25) Alexander Friedmann proposed expanding universe; (26) Friedmann–Lemaître–Robertson–Walker metric cosmological model; (27) Stern–Gerlach experiment; (28) Galaxies discovered; (29) Particle nature of photons confirmed by observation of photon momentum; (30) Bose–Einstein statistics; (31) De Broglie wave; (32) Matrix mechanics; (33) Niels Bohr & Max Planck: Quantum mechanics; (34) Stellar structure understood; (35) Fermi-Dirac Statistics; (36) Schrödinger Equation; (37) Uncertainty principle; Big Bang; (37) Dirac equation; (38) Max Born interpretation of the Schrödinger equation; (38) Paul Dirac proposes the antiparticle; (39); Expansion of the universe confirmed; (40) Antimatter discovered; (41) Neutron discovered; (42) Invention of the electron microscope; (43) Chandrasekhar limit for black hole collapse; (44) Muon discovered; (45) Superfluidity discovered; (46) Nuclear fission discovered; (47) Stellar fusion explains energy production in stars; (48) Uranium fission discovered; (49) Feynman path integral; (50) Theory of magnetism in 2D: Ising model; (51) Pion discovered; (52) Quantum electrodynamics; (53) Invention of the maser and laser; (54) Feynman diagrams; (55) Electron neutrino discovered.

*(1951-1999)*. (56) Parity violation proved; (57) BCS theory explaining superconductivity; (58) Role of topology in quantum physics predicted and confirmed; (59) SU(3) theory of strong interactions; (60) Muon neutrino discovered; (61) Confirmed the conserved vector current theory for weak interactions; (61) Quarks predicted; (62) Bell's Theorem initiates quantitative study of quantum entanglement; (63) Unification of weak interaction and electromagnetism (electroweak theory); (64) Solar neutrino problem found; (65) Pulsars (rotating neutron stars) discovered; (66) Experimental evidence for quarks found; (67) Dark matter theories; (68) Standard Model of elementary particles invented; (69) Helium 3 superfluidity; (70) Renormalization group; (71) Black Hole Entropy; (72) Black hole radiation (Hawking radiation) predicted; (73) Charmed quark discovered; (74) Tau lepton found; (75) Bottom quark found; (76) Strangeness as a signature of quark-gluon plasma predicted; (77) Richard Feynman proposes quantum computing; (78) Quantum Hall effect; (79) IanGuth Theory of cosmic inflation proposed; (80) Fractional quantum Hall effect discovered; (81) W and Z bosons directly observed; (82) First laboratory implementation of quantum cryptography; (83) Quantum teleportation of unknown states proposed; (84) Shor's algorithm discovered, initiating the serious study of quantum computation; (85) Matrix models/M-theory; (86) Bose–Einstein condensate observed; (87) Top quark discovered; (88) Accelerating expansion of the universe discovered by the Supernova Cosmology Project and the High-Z Supernova Search Team; (89) Atmospheric neutrino oscillation established; (90) Slow light experimentally demonstrated.

*(2000-2020)*. (90) Quark-gluon plasma found; (91) Tau neutrino found; (92) Solar neutrino oscillation was observed which resolved the solar neutrino problem; (93) WMAP observations of cosmic microwave background; (94) Isolation and characterization of graphene; (95) 16-year study of stellar orbits around Sagittarius_A* provide strong evidence for a supermassive black hole at the centre of the Milky Way galaxy; (96) Planck (a European Space Agency space-based observatory) began the observations of cosmic microwave background; (97) Higgs boson found by the Compact Muon Solenoid and ATLAS experiments at the Large Hadron Collider; (98); Gravitational waves are observed; (99) First image of a Black hole and (100) The first room-temperature superconductor identified.

This list of significant discoveries helps us to look atdifferent periods in physics that resulted in the crisis afterwards. We can: (1) detect an upward movement of modern physics in the second half of the 19th century which decelerated at the end of the century; (ii) localize the qualitative jump of modern physics in the first half of the 20th century and eventually (iii) the occurrence of the crisis. In the following text, all definitions of the crisis of physics, including ours, will be referred to as the crisis in physics.

What have caused and created this crisis? It seems that the crisis in physics has at least three aspects: psychological, epistemological as well as social and economic aspects.

(1) Psy*chological aspect*. Let us make a digression regarding the term *escape* in the context of positive freedom which was defined by Fromm (1941). Based on the Fromm's definition, it is the capacity for "spontaneous relationship to man and nature, a relationship that connects the individual with the world without eliminating his individuality". Freedom is also accompanied by loneness as well as an inability to exert individual power and "we use several different techniques to alleviate the anxiety associated with our *perception of freedom*, including automaton, conformity, authoritarianism, destructiveness, and individuation." (*emphasis added* by the authors). The most common of these mechanisms is *conformity*. Fromm stated that people conform to larger society and gain power and a sense of belonging by behaving similarly. This power of the masses assists us in not feeling lonely and helpless, but unfortunately, it removes our individuality. How is this phenomenon connected with the crisis in physics? Specifically, the escape from *science* could be seen as an analogy to the escape from *freedom* or "escape from freedom of choice of the research subject". How is it possible for physicists to escape from physics today? Physicists are willing to believe that cause can be determined, and they accept an obvious solution and declare it *a right* explanation. On the contrary, they avoid a random event which almost always affects and changes physical systems. Why? Because random events evade a clear explanation or rules which they find nearly impossible to deal with. The ignorance of chance and random events in physical processes leads to *scientific conformity* and the creation of *mainstream* in science (hence physics) or the mechanism of escape from physics. Note that the syntagma "mainstream" refers to physicists who have adopted attitudes that are opposite to the Anderson's understanding of physics. Finally, our mind is filled with causality, and we primarily say *because* instead of accepting a random event.

(2) *Epistemological aspect.* (i) *Limits of the precision of certainty and strength of Gödel's theorem*. There is no doubt that physics is approaching some limits, the only question is whether these are the limits to how far physics can reach or its end? To clarify our position, we will rely on Gödel's incompleteness theorems. Barrow (2006) considered some informal aspects of these theorems and their underlying assumptions and discussed some of the responses to these theorems by those seeking to draw conclusions from them about the completability of theories of physics. In the same paper he enhanced "that there is no reason to expect Gödel incompleteness to handicap the search for a description of the laws of Nature, but we do expect it to limit what we can predict about the outcomes of those laws". Stanley Jaki believed that Gödel's theorem prevented us from gaining understanding of the cosmos as a necessary truth: "Clearly then no scientific cosmology, which of necessity must be highly mathematical, can have its proof of consistency within itself as far as mathematics goes. In the absence of such consistency, all mathematical models, all theories of elementary particles, including the theory of truth that the world can only be what it is and nothing else. This is true even if the theory happened to account with perfect accuracy for all phenomena of the physical world known at a particular time." (Jaki, 1980). It constitutes a fundamental barrier to the understanding of the universe: "It seems that on the strength of Gödel's theorem that the ultimate foundations of the bold symbolic constructions of mathematical physics will remain embedded forever in that deeper level of thinking characterized both by the wisdom and by the haziness of analogies and intuitions. For the speculative physicist this implies that there are limits to the precision of certainty, that even in the pure thinking of theoretical physics there is a boundary...An integral part of this boundary is *the scientist himself, as a thinker*..." (Jaki, 1966); (*emphasis added* by the authors). (ii) *Limits of decoding information*. Advances in information theory thought to be of critical importance in modern physics. For instance, they are important for detecting the gravitational waves provided by the LIGO (Laser Interferometer Gravitational-Wave Observatory) (Kovalsky and Hnilo 2021). However, there are some limitations and even if we move the borders of cognition in physics, it remains the problem of decoding information. Each meaningful information is encoded in patterns which have different complexity and is always less complex than maximum complexity (randomness). Unlike easily observed and immediately decodable patterns (e.g. visual input, native language), our cognitive capacity struggles extracting information from noise. Furthermore, the presence of a pattern (*structural information*) in noise is a necessary but not a sufficient condition for the presence of meaning (*semantic information*). Besides, there are other aspects that should be considered: (i) how we interpret information and (ii) can we search for information outside established concepts? Namely, people are not used to thinking abstractly (Taleb 2016). In other words, they need a proven set of rules and facts as a basis for further induction. Consequently, most scientists are focused on and seek the confirmation of what happened and ignore what could have happened. We lack the dimension of abstract thinking, and we are not even aware of that. Therefore, physicists initiate "the trap of induction "and search for information which is easy to comprehend and not information which is hardly reachable to our minds.

It is worth mentioning is the effect of scientists' emotions and intuition on science and scientific work. Intuition is the accumulation of experiences which are assimilated unconsciously (Damasio 1994). Einstein wrote an interesting consideration on the importance of the intuition in an essay (1919): "In these unstable times, I present to the reader this small, objective but passionate reflection …For you never know if, even in the future, you will not find an experience that contradicts its consequences; and, still, one can always conceive of other systems of thought capable of connecting the same facts given. If two theories are available, both compatible with the given factual material, then there is no other criterion to prefer one or the other, other than the intuitive view of the researcher. Thus, we can understand how insightful researchers, who dominate theories and facts, can still be passionate advocates of contradictory theories." Emotions are essential in science including the justification as well as the discovery of hypotheses (Thagard 2002). Note that Gödel's work has had a huge impact on speculations about the limitations of the human mind (Barrow 2006).

(3) *Social and economic aspects.* The fact that a major breakthrough in physics has not recently appeared influences the general social attitude towards research. Building large machines (colliders, telescopes) cannot be justified neither by the pure curiosity, nor promising results, but mostly, just as a cosmic flight or Formula 1 racing contribute to the technology, by immediate applications in everyday life. This reasoning affects decision makers who deal with research planning and financing which resulted in misinterpretations of scientific results just to make them look like something applicable. (Some might remember the stories of the "Negative mass" implying antigravity, but actually it was "negative effective mass", a phenomenon well known in condensed matter physics, implying that on the plot of energy vs. momentum (wave-vector), the slope is opposite to the usual one). This aspect is sometimes forgotten, but when it comes to experimental research, it is of great importance.

*How do we see the "ending" of physics?* It appears that a lot of things have come to an end lately. First, Francis Fukuyama, a political scientist, proclaimed *The End of History and the Last Man*. Then David Lindley announced *The End of Physics* in which he argued that the theory of everything derived from particle physics will be full of untested and untestable assumptions. If physics is the source of such speculation, it will eventually detach from science and become modern mythology. *This will be the end of physics as we know it* (*emphasis added* by the authors). Finally, John Horgan, a science journalist, went beyond Lindley's opinions to include all sciences. His book consists of interviews with approximately 45 prominent scientists. He organized those interviews into chapters, each containing the discussion about the end of one field: philosophy, physics, cosmology, evolutionary biology, social science, neuroscience, "chaoplexity", "limitology", and machine science. We can notice that Fukuyama did not anticipate some further developments of civilization. He did not recognize the possibility of the appearance of the *Black Swan*, which caused tectonic changes (the outbreak of COVID-19 (see subchapter 1.1)), but other observations on "the end of science" are correct. Indeed, all big problems in physics have been solved, or soon will be. Apparently, there are not more truly fundamental discoveries, such as quantum mechanics, relativity, the Big Bang and beyond (cosmology), left to be made. Therefore, some possibilities for future trends in physics are: (i) it will be aimed at less interesting activities, the research on detailed implications of basic theories, applied problems, etc. and (ii) it will be less stimulating and attractive to *stellar* scientists who are able to make vertical progress.

The Nobel Laureate Philip Anderson presented his observations on physics and its future progression and directions towards fundamental discoveries (Anderson, 1972). That was approximately 35-40 after the most fundamental discoveries in physics were made. At the beginning of the article Anderson wrote: "In physics there are two clearly distinguished trends" which he named "intensive" and "extensive" research. Intensive research includes the fundamental laws while extensive research comprises the explanation of phenomena in terms of established fundamental laws. As stated by him: "There are two dimensions to basic research. The frontier of science extends all along a long line from the newest and most modern intensive research, over the extensive research recently spawned by the intensive research of yesterday, to the broad and well-developed web of extensive research activities based on intensive research of past decades." He pointed out the nature of the hierarchical structure of science and physics. At each level of complexity entirely new properties appear, and the understanding of new behavior requires research which is fundamental in its nature. Anderson described his thoughts on fundamental science through his experience: "The effectiveness of this message (about intensive and extensive sciences, (*comment by* the authors)) may be indicated by the fact that I heard it quoted recently by a leader in the field of materials science, who urged the participants at a meeting dedicated to "fundamental problems in condensed matter physics" to accept that there were few or no such problems and that nothing was left but extensive sciences which he seemed to *equate* with *device engineering* (*emphasis added by* the authors)". His comments can be interpreted as the strong criticism of efforts at replacing science with technology or, even as an attempt to transform philosophy into science; philosophy is not science since it employs rational and logical analysis as well as conceptual clarification while science employs empirical measurements. Similarly, the technology cannot replace the science. In our opinion, if we consider his conclusions, we can notice continuity and not the end of physics. Our perspective on future developments is similar to Dejan Stojaković's: "Further progress of physics will require a synthesis of collective efforts in seemingly unrelated fields - physics, mathematics, chemistry, biology, medicine, psychology, and even arts such as literature, painting and music" (Stojković, 2020). An additional comment could clarify this sentence more precisely. *Physics could contribute to the domain of fundamental discoveries analogously to the way entities in a complex system contribute to a whole system.*